\documentclass[12pt]{article}
\usepackage{fullpage}

\usepackage{theorem}
\newtheorem{theorem}{Theorem}[section]
\newtheorem{proposition}[theorem]{Proposition}
\theorembodyfont{\normalfont}
\newtheorem{definition}[theorem]{Definition}
\newtheorem{example}[theorem]{Example}
\newtheorem{remark}[theorem]{Remark}

\author{Dominique Duval \\ \small{
LJK, Universit\'e de Grenoble, France, \url{Dominique.Duval@imag.fr}} 
}
\title{Diagrammatic Inference}

\usepackage{amssymb,amsmath} 
\usepackage[all]{xy} 
\usepackage{url}

\newcommand{\ie}{\textit{i.e.}}
\newcommand{\resp}{\textit{resp.\ }}

\newcommand{\Rea}{\mathbf{Real}}   
\newcommand{\Proto}{\mathbf{Proto}}   
\newcommand{\Type}{\mathbf{Type}}   

\newcommand{\Mod}{\mathrm{Mod}} 

\newcommand{\id}{\mathrm{id}}     
\newcommand{\Id}{\mathrm{Id}}     
\newcommand{\Lim}{\mathrm{Lim}}
\newcommand{\op}{\mathrm{op}} 

\newcommand{\gr}{\mathrm{Gr}}

\newcommand{\fp}{\mathrm{Fp}} 
\newcommand{\fpcat}{\mathrm{FpCat}} 
\newcommand{\llim}{\mathrm{Lim}} 
\newcommand{\fpsk}{\mathrm{FpSk}}

\newcommand{\effgr}{\mathrm{EGr}}

\newcommand{\fpeffcat}{\mathrm{FpECat}} 
\newcommand{\fpeffsk}{\mathrm{FpESk}} 

\newcommand{\eq}{\mathrm{Eq}}   
\newcommand{\effeq}{\mathrm{EEq}}  

\newcommand{\HH}{\mathcal{H}} 
\newcommand{\CC}{\mathcal{C}} 

\newcommand{\fr}[2]{{#1}\backslash{#2}}
\newcommand{\frsyn}[2]{{#1}/{#2}}

\newcommand{\Lto}{\longrightarrow} 
\newcommand{\lto}{\leftarrow}  
 
\newcommand{\upto}[1]{\stackrel{#1}{\rightarrow}}  
\newcommand{\lupto}[1]{\stackrel{#1}{\leftarrow}}  
\newcommand{\iso}{\cong}    
\newcommand{\To}{\Rightarrow}   
\newcommand{\adjto}{\rightleftarrows} 
\newcommand{\ento}{\stackrel{\dashleftarrow}{\rightarrow}}  
\newcommand{\lento}{\stackrel{\dashrightarrow}{\leftarrow}}  
\newcommand{\opto}{\Lto\hspace{-18pt}\times\hspace{10pt}}  

\newcommand{\grJ}{\mathbf{J}}  
\newcommand{\catA}{\mathbf{A}} 
\newcommand{\catC}{\mathbf{C}} 
\newcommand{\catD}{\mathbf{D}} 
\newcommand{\catV}{\mathbf{V}}  
\newcommand{\skE}{\mathbf{E}}   

\newcommand{\catS}{\mathbf{S}} 
\newcommand{\catSsim}{\mathbf{S1}} 
\newcommand{\catShat}{\mathbf{S2}} 
\newcommand{\catT}{\mathbf{T}} 

\newcommand{\Set}{\mathbf{Set}} 
\newcommand{\Gr}{\mathbf{Gr}} 
\newcommand{\Cat}{\mathbf{Cat}} 
\newcommand{\FpCat}{\mathbf{FpCat}} 
\newcommand{\LCat}{\mathbf{LCat}} 
\newcommand{\FpSk}{\mathbf{FpSk}}   
\newcommand{\LSk}{\mathbf{LSk}}   
\newcommand{\DiaLog}{\mathbf{DiaLog}} 

\newcommand{\funG}{G} 
\newcommand{\funU}{U} 
\newcommand{\funUsim}{U_1} 
\newcommand{\funUhat}{U_2} 
\newcommand{\funF}{F} 
\newcommand{\funFsim}{F_1} 
\newcommand{\funFhat}{F_2} 
\newcommand{\funL}{L} 
\newcommand{\funI}{I} 
\newcommand{\funQ}{Q} 
\newcommand{\funC}{C} 
\newcommand{\funY}{\mathcal{Y}} 

\newcommand{\logL}{\mathcal{L}} 
\newcommand{\logR}{\mathcal{R}} 

\newcommand{\eqwe}{\lesssim}
\newcommand{\lr}[1]{\langle #1 \rangle}
\newcommand{\Spec}{\mathrm{Spec}}   
\newcommand{\setS}{\mathbb{S}}   

\newcommand{\deco}{\mathrm{dec}}   
\newcommand{\expl}{\mathrm{exp}}   
\newcommand{\simp}{\mathrm{sim}}   

\begin{document}

\maketitle

\begin{abstract}
Diagrammatic logics were introduced in 2002,
with emphasis on the notions of specifications and models.
In this paper we improve the description of the inference process,
which is seen as a Yoneda functor on a bicategory of fractions.
A diagrammatic logic is defined from a morphism of limit sketches
(called a propagator) which gives rise to an adjunction, 
which in turn determines a bicategory of fractions. 
The propagator, the adjunction and the bicategory 
provide respectively the syntax, the models and the inference process for the logic.
Then diagrammatic logics and their morphisms are applied to the semantics 
of  side effects in computer languages. 
\end{abstract}

\section{Introduction} 
\label{sec:int}

The framework of \emph{diagrammatic logics} was introduced in \cite{Du03}, 
after \cite{DL02}. 
It relies on well-known categorical notions like 
Ehresmann's sketches \cite{Eh68} and 
Gabriel and Zisman's categories of fractions \cite{GZ67}.
Diagrammatic logics have been influenced 
by Lair's ``\emph{trames}'' \cite{Lai87} and 
by Makkai's \emph{sketch entailments} \cite{Mak97}.
They share many common features with 
other approaches of categorical logic, among which 
\cite{Lam68,Law69a,Law69b,Mi75,Se79,BW94,Bj05,Gu07}.
While categorical logic traditionally relies on viewing logical theories as categories,
or sometimes as 2-categories, 
diagrammatic logic allows more general kinds of theories.
This is motivated by applications to computer science, 
like the application to side-effects that is presented at the end of the paper. 

Diagrammatic logics were introduced 
in order to deal with some unusal kinds of logics 
and with morphisms between  logics. 
They can be used for proving properties 
of computational languages with effects, in a natural and powerful way, 
and for providing a notion of model for these languages,
with guaranteed soundness properties 
\cite{DR06,DDLR06,DDR07}.
A diagrammatic logic $\logL$ is defined from a special kind of \emph{adjunction}, 
which itself comes from a special kind of \emph{morphism of limit sketches}.
The adjunction provides the \emph{models} and the \emph{inference process}
for the logic $\logL$,
while the limit sketches and their morphism provide a \emph{syntax} for~$\logL$.
In this paper, the inference process for a diagrammatic logic 
is defined as a \emph{Yoneda functor on a bicategory}.
Bicategories, as introduced by B\'enabou \cite{Be67},
did not appear explicitly in \cite{Du03}.
In this paper, we show that they play a major role in clarifying 
the notion of diagrammatic inference.

We define a \emph{logical adjunction} $\funF\dashv\funU:\catS\adjto\catT$
(where $\funF:\catS\to\catT$ and $\funU:\catT\to\catS$) 
as an adjunction where $\catS$ is cocomplete and $\funU$ is full and faithful.
Then $\catS$ is called the category of \emph{specifications}
and $\catT$ the category  of \emph{theories}.
Typically, a specification $\Sigma$ is a collection of axioms 
and a theory $\Theta$ is a collection of theorems that is closed under inference.
The \emph{models} of $\Sigma$ with values in $\Theta$
are defined from the bijection 
$ \catS(\Sigma,\funU\Theta) \iso \catT(\funF\Sigma,\Theta) $ 
(natural in $\Sigma$ and $\Theta$) 
in section~\ref{sec:mod}, as in \cite{Du03}.

For a given logical adjunction $\funF\dashv\funU:\catS\adjto\catT$,
the aim of inference is to determine, for a given specification $\Sigma$,
some \emph{generalized elements} of $\funF\Sigma$,
which means, some morphisms in $\catT$ with codomain $\funF\Sigma$.
Let $\catSsim$ be the category of \emph{classes of fractions} 
$\fr{\tau}{\sigma}$ of $\catS$, 
with numerator any morphism $\sigma$ in $\catS$ and 
with denominator an \emph{entailment} $\tau$, \ie, 
a morphism in $\catS$ such that $\funF\tau$ is invertible in $\catT$.
Then there is an equivalence $\funFsim\sim\funUsim:\catSsim\adjto\catT$ 
with $\funFsim\Sigma=\funF\Sigma$, 
so that there is a bijection
$ \catSsim(\funUsim\Theta,\Sigma) \iso \catT(\Theta,\funF\Sigma) $,
natural in $\Theta$ (in $\catT$) and $\Sigma$ (in $\catSsim$).
The category $\catSsim$ of classes of fractions is obtained from 
the \emph{bicategory of fractions} $\catShat$.
Let $\funC$ denote the \emph{connectivity functor} from $\Cat$ to $\Set$, 
which identifies the connected objects in every category.
Functors $\funUhat:\catShat\to\catT$ 
and $\funFhat:\catT\to\catShat$ 
are defined, with $\funFhat\Sigma=\funF\Sigma$, such that
the previous bijection can be stated, more precisely, as 
$\funC(\catShat(\funUhat\Theta,\Sigma)) \iso \catT(\Theta,\funF\Sigma)$, 
naturally in $\Theta$ (in $\catT$) and $\Sigma$ (in $\catShat$).
This is the key for defining the inference process: 
the  \emph{inference rules} are fractions, 
from which all \emph{proofs} can be derived, 
an \emph{inference step} is defined as a composition in $\catShat$, 
so that the \emph{inference process}
can be seen as the Yoneda covariant functor on the bicategory $\catShat$. 
It should be noted that the composition in $\catShat$, 
for each inference step, requires a pushout in $\catS$. 
Inference is studied, along these lines, in section~\ref{sec:ded}.

A \emph{limit sketch} $\skE$ 
is made of a graph together with some potential identities, composites and limits, 
which turn $\skE$ into a generator for a complete category. 
Here, as in \cite{Du03}, morphisms of limit sketches are called \emph{propagators}.
Each limit sketch $\skE$ gives rise to a category $\Rea(\skE)$ of realizations,
or ``loose models'', which is cocomplete. 
It is known from \cite{Eh68} 
that each propagator $P:\skE_S\to\skE_T$ determines  
an adjunction $\funF\dashv\funU:\catS\adjto\catT$,  
where $\catS=\Rea(\skE_S)$ and $\catT=\Rea(\skE_T)$.
According to \cite{Du03}, any propagator $P$ can be modified 
in a reasonable way (reminded in theorem~\ref{thm:syn-dec}) in order to get~$\funU$ full and faithful.
Then $P$ is called a \emph{logical propagator},
and $\funF\dashv\funU$ is a logical adjunction.
A \emph{diagrammatic logic} $\logL$ 
is defined as an equivalence class of logical propagators.
An \emph{inference system} for $\logL$ is a propagator in the class $\logL$  
which consists of adding inverses to some arrows,
it provides the inference rules. 
This is studied in section~\ref{sec:syn}, 
where in addition it is checked that an inference system may satisfy 
relevant \emph{finiteness conditions} to be called a \emph{syntax}.

The paper ends up, in section~\ref{sec:exa}, with an application to side-effects 
in computer languages, which provides a categorical base for \cite{DDR07}. 

Some familiarity with category theory is assumed:  
most of it can be found in \cite{Mac98},
and in \cite{Le98} for bicategories. 
We use the category $\Set$ of sets, the 2-category $\Cat$ of categories, 
and so on, without mentioning the size issues.
Sometimes we use the symbol ``$\opto$'' for contravariant functors, 
so that $\catC\opto\catD$ means either $\catC^{\op}\to\catD$ or $\catC\to\catD^{\op}$, 
and we denote by $\catD^{\catC,\op}$ the category of contravariant 
functors from $\catC$ to $\catD$. 
For an introduction to the theory of sketches (``\emph{esquisses}'', in french),
see \cite{CL84,CL88,BW99}, and \cite{We93} for additional references.
There are many kinds of sketches (linear sketches, 
finite product sketches, limit sketches,\dots), 
which correspond to different kinds of logic.
In addition, there are many variants for each kind of sketches,
however the choice of some variant only matters for the technical details.

\section{Models} 
\label{sec:mod}

\subsection{Adjunctions}
\label{subsec:mod-adj}

An \emph{adjunction}
is a pair of functors $\funF:\catS\to\catT$ (\emph{the left adjoint}) 
and $\funU:\catT\to\catS$ (\emph{the right adjoint})
together with a bijection, 
natural in $\Sigma$ (in $\catS$) and $\Theta$ (in $\catT$):
\begin{equation}
\label{eqn:mod}
  \catS(\Sigma,\funU\Theta) \iso \catT(\funF\Sigma,\Theta) 
\end{equation}
This is denoted $\funF\dashv\funU$,
or more precisely $\funF\dashv\funU:\catS\adjto\catT$.
An adjunction defines two natural transformations, 
the \emph{unit} $\eta:\Id_{\catS}\To \funU\funF$ 
and the \emph{counit} $\varepsilon:\funF\funU\To\Id_{\catT}$.
When both $\eta$ and $\varepsilon$ are  natural isomorphisms,
the adjunction is called an \emph{equivalence (of categories)},
which is denoted $\funF\sim\funU:\catS\adjto\catT$.
A \emph{morphism of adjunctions},
from $\funF\dashv\funU:\catS\adjto\catT$ to $\funF'\dashv\funU':\catS'\adjto\catT'$, 
is a pair of adjunctions $\funF_S\dashv\funU_S:\catS\adjto\catS'$
and $\funF_T\dashv\funU_T:\catT\adjto\catT'$ 
such that $\funU\circ\funU_T=\funU_S\circ\funU'$,
from which follows a natural isomorphism $\funF_T\circ\funF\iso\funF'\circ\funF_S$.
A morphism made of two equivalences is an \emph{equivalence of adjunctions}.

\subsection{Logical adjunctions}
\label{subsec:mod-logic}

\begin{definition}
\label{defi:mod-log-adj} 
A \emph{logical adjunction} is an adjunction $\funF\dashv\funU:\catS\adjto\catT$
such that the category $\catS$ is cocomplete and the functor $\funU$ is full and faithful.
For instance, a \emph{full reflection} between cocomplete categories 
is a logical adjunction. 
\end{definition}

From now on in section~\ref{sec:mod}, 
a logical adjunction $\funF\dashv\funU:\catS\adjto\catT$ is chosen.
\begin{definition}
\label{defi:mod-spth}
The category of \emph{specifications} and the category of \emph{theories}
are $\catS$ and $\catT$, respectively.
For each specification $\Sigma$ and theory~$\Theta$,
the set of \emph{models} of $\Sigma$ with \emph{values} in $\Theta$ is
$\Mod(\Sigma,\Theta) = \catT(\funF\Sigma,\Theta)$,
so that $\Mod(\Sigma,\Theta) \iso \catS(\Sigma,\funU\Theta)$.
\end{definition}
This gives rise to the functor $\Mod:\catS\times\catT^{\op}\to\Set$. 
It may happen that this functor takes its values in $\Cat$.
The next result is a direct consequence of adjunction, it will be used 
in section~\ref{sec:exa}.
\begin{proposition}
\label{prop:mod-mod}
Let us consider a morphism 
$\langle \funF_S\dashv\funU_S,\funF_T\dashv\funU_T\rangle$ 
from $\funF\dashv\funU$ to a logical adjunction 
$\funF'\dashv\funU':\catS'\adjto\catT'$. 
Then there is a bijection, natural in 
$\Sigma$ (in $\catS$) and $\Theta'$ (in $\catT'$):
   $$ \Mod_{\funF\dashv\funU}(\Sigma,\funU_T\Theta') \iso
   \Mod_{\funF'\dashv\funU'}(\funF_S\Sigma,\Theta') \;.$$
\end{proposition}

\begin{example}[Equational logic]
\label{exa:mod-equ}
Let $\Gr$ denote the category of graphs 
and $\FpCat$ the category of categories with chosen \textbf{F}inite \textbf{P}roducts.
The inclusion of $\FpCat$ in $\Gr$ gives rise to a reflection, that is not full. 
The finite product sketches are defined now, 
as a kind of intermediate notion between graphs and categories
with chosen finite products.
First, a \emph{linear sketch} $\skE$ is a graph where
for some points $X$ 
    there is a loop $\id_X:X\to X$
    called \emph{the (potential) identity of $X$}, 
for some consecutive arrows $f:X\to Y$, $g:Y\to Z$ 
    there is an arrow $g\circ f:X\to Z$  
    called \emph{the (potential) composite of $f$ and $g$}. 
Then, a \emph{finite product sketch} $\skE$ is a linear sketch where
for some finite families of points $(X_1,\dots,X_k)$ (with $k\geq0$)
    there is a discrete cone $(p_j:\prod_{i=1}^nX_i\to X_j)_{1\leq j\leq k}$ in $\skE$ 
    called \emph{the (potential) product} of $X_1,\dots,X_k$). 
No additional axiom has to be satisfied.
A \emph{morphism of finite product sketches}
is a morphism of graphs which preserves all potential features.
This yields the category $\FpSk$ of finite product sketches.
The variants for this definition include: 
a potential identity for each point, 
and/or a potential composite for each pair of consecutive arrows,
or ``diagrams'' instead of composites, 
and/or any number of potential products (often called 
\emph{distinguished cones}) for each finite discrete base. 
A category with chosen finite products can be seen as a finite product sketch,
with its chosen products as potential products; 
this inclusion of $\FpCat$ in $\FpSk$ gives rise to a full reflection.
Since the category $\FpSk$ is cocomplete, we get a logical adjunction
$\funF_{\fp}\dashv\funU_{\fp}:\FpSk\adjto\FpCat$. 
The category of sets, with some choice for the finite products of sets,
defines a theory with respect to this logical adjunction. 
Every equational specification $\Spec$
can be seen as a finite product sketch $\Sigma$:
the sorts, operations and equations become points, 
arrows and equalities of arrows, respectively \cite{BW99}.  
The diagrammatic models of $\Sigma$ with values in the theory of sets 
can be identified with models of $\Spec$, which can be called
the ``strict'' models of $\Spec$.
\end{example}

\begin{example}[Limit logic]
\label{exa:mod-lim}
It is easy to generalize example~\ref{exa:mod-equ} by replacing 
finite products with limits.
The resulting logic will be called the \emph{limit logic}.
Let $\LCat$ denote the category of categories with chosen \textbf{L}imits.
We define \emph{limit sketches} as
an intermediate notion between graphs and categories with chosen limits; 
they are similar to \emph{projective sketches} \cite{CL84}.
A \emph{limit sketch} $\skE$ is a linear sketch where
for some diagrams $b:\grJ\to\skE$ 
    there is a commutative cone $(p_J:\Lim(b)\to X_J)_J$ with base $b$ in $\skE$ 
    (\ie, the $J$'s are the points of $\grJ$
    and $b(j)\circ p_J=p_K$ for each arrow $j:J\to K$ in $\grJ$), 
    called \emph{the (potential) limit of $b$}.
No additional axiom has to be satisfied.
A \emph{morphism of limit sketches}
is a morphism of graphs which preserves all potential features.
This yields the category $\LSk$ of limit sketches.
There exist also many variants for this definition. 
The inclusion of $\LCat$ in $\Gr$ gives rise to a reflection, that is not full. 
A category with chosen limits can be seen as a limit sketch,
with its chosen limits as potential limits;
this inclusion of $\LCat$ in $\LSk$ gives rise to a full reflection.
Since the category $\LSk$ is cocomplete, we get a logical adjunction
$\funF_{\llim}\dashv\funU_{\llim}:\LSk\adjto\LCat$. 
For every limit sketch $\Sigma$, a diagrammatic model of $\Sigma$
with values in the category of sets, with some choice for the limits of sets,
maps the points of $\Sigma$ to sets, 
its arrows to functions, and its potential limits to the chosen limits: 
so, the diagrammatic models are ``strict'' models.
\end{example}

Examples~\ref{exa:mod-equ} and~\ref{exa:mod-lim} are easily generalized 
to other kinds of sketches and categories with structure. 
Quoting \cite{We93}, following Lawvere \cite{Law69b} 
(in this paper we say ``kind'' instead of ``type''): 
\textsl{``Let $E$ be a type of sketch, 
determined by what sorts of cones and cocones are allowed in the sketch. [\dots] 
Corresponding to each type $E$ there is a type of category,
required to have all limits, respectively colimits,
of the type of cones, respectively cocones, allowed by $E$. 
Likewise, there is a type of functor,
required to preserve that type of limits or colimits. [\dots]
Given a type $E$, we will refer to $E$-sketches, $E$-categories and $E$-functors. 
Following Lawvere, we will refer to $E$ as a doctrine. [\dots]
A doctrine can be a type of category requiring other structure 
besides limits and colimits (precisely, any type of category 
definable essentially algebraically over the category of categories).''}
Moreover, in section~\ref{sec:exa}, we define a logical adjunction 
such that its theories are not defined as ``categories with structure''.

\section{Inference}
\label{sec:ded}

\subsection{Fractions}
\label{subsec:ded-frac}

This section relies on chapter~1 of \cite{GZ67}.
We insist on the fact that fractions form the objects of a  \emph{bicategory} 
\cite{Be67,Le98}.
Let $\catS$ be a cocomplete category 
and $\funF:\catS\to\catT$ a functor preserving colimits, 
which is satisfied whenever there is a logical adjunction 
$\funF\dashv\funU:\catS\adjto\catT$.

A morphism $\tau:\Sigma\to\Sigma'$ in $\catS$ is a \emph{$\funF$-isomorphism}
if $\funF\tau$ is an isomorphism in $\catT$;
this is denoted $\tau:\Sigma\ento\Sigma'$. 
A \emph{(left $\funF$-)fraction} from $\Sigma$ to $\Sigma_1$
is a cospan $\fr{\tau}{\sigma}:\Sigma\to\Sigma'_1\lento\Sigma_1$
where $\sigma:\Sigma\to\Sigma'_1$ is any morphism in $\catS$
and $\tau:\Sigma_1\ento\Sigma'_1$ is a $\funF$-isomorphism.
A \emph{morphism of fractions}  from $\Sigma$ to $\Sigma_1$, 
say $\alpha:\fr{\tau_1}{\sigma_1}\To\fr{\tau_2}{\sigma_2}$ 
where $\fr{\tau_1}{\sigma_1}:\Sigma\to\Sigma'_1\lento\Sigma_1$ 
and $\fr{\tau_2}{\sigma_2}:\Sigma\to\Sigma'_2\lento\Sigma_1$, 
is a morphism $\alpha:\Sigma'_1\to\Sigma'_2$ in $\catS$ 
such that $\alpha\circ\sigma_1=\sigma_2$ 
and $\alpha\circ\tau_1=\tau_2$.
This last equality implies that $\alpha$ is a $\funF$-isomorphism.
The fractions from $\Sigma$ to $\Sigma_1$ together with their morphisms
form a category $\catShat(\Sigma,\Sigma_1)$.

The \emph{composite} of two consecutive fractions
$\fr{\tau_1}{\sigma_1}:\Sigma_0\to\Sigma'_1\lento\Sigma_1$ 
and $\fr{\tau_2}{\sigma_2}:\Sigma_1\to\Sigma'_2\lento\Sigma_2$
is the fraction 
$(\fr{\tau_2}{\sigma_2})\circ(\fr{\tau_1}{\sigma_1}):
\Sigma_0\to\Sigma''_2\lento\Sigma_2$
with numerator $\sigma=\sigma'\circ\sigma_1$ 
and denominator $\tau=\tau'\circ\tau_2$, 
where $\sigma'$ and $\tau'$ are obtained from the pushout 
of $\tau_1$ and $\sigma_2$.
Since $\funF$ preserves colimits,
the properties of pushouts prove that $\tau'$ is a $\funF$-isomorphism,
so that $\tau$ also is a $\funF$-isomorphism:
$$
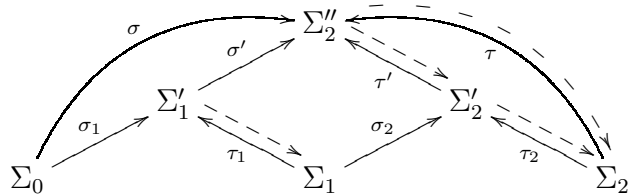
\xymatrix@R=1pc@C=3pc{
&& \Sigma''_2 \ar@<1ex>@{-->}[rd]  \ar@/^5ex/@<1ex>@{-->}[rrdd] && \\
&\Sigma'_1 \ar[ru]^{\sigma'} \ar@<1ex>@{-->}[rd] & & 
   \Sigma'_2  \ar[lu]^{\tau'} \ar@<1ex>@{-->}[rd] &\\
 \Sigma_0 \ar[ru]^{\sigma_1}  \ar@/^5ex/[rruu]^{\sigma} && 
   \Sigma_1 \ar[lu]^{\tau_1} \ar[ru]^{\sigma_2} &&
   \Sigma_2 \ar[lu]^{\tau_2}  \ar@/_5ex/[lluu]^{\tau} \\
}$$
Together with the identities $\fr{\id}{\id}$,
this forms the \emph{bicategory of fractions} $\catShat$:
it has the same objects as $\catS$, 
the fractions as morphisms (or 1-cells),
and the morphisms of fractions as 2-cells.
Every morphism $\sigma:\Sigma\to\Sigma_1$ in $\catS$ can be identified to 
the fraction $\fr{\id_{\Sigma_1}}{\sigma}$, 
so that $\catS$ is a wide subcategory of the bicategory $\catShat$ 
(\emph{wide} means they have the same objects); 
the inclusion functor is denoted $\funI:\catS\to\catShat$. 

Let $b:\grJ\to\catS$ be a diagram in $\catS$.
A \emph{lax commutative cocone} in $\catShat$ with base $b$
is a cocone $(\fr{\tau_J}{\sigma_J}:b(J)\to\Sigma)_J$, for all objects $J$ in $\grJ$,  
together with a 2-cell 
$\alpha_j:\fr{\tau_J}{\sigma_J}\To\fr{\tau_K}{\sigma_K}\circ b(j)$
for each arrow $j:J\to K$ in $\grJ$.
A cocone $(h_J:b(J)\to\HH)_J$ in $\catS$ with base $b$
is a \emph{lax colimit} in $\catShat$ 
if it is commutative and if for each lax commutative cocone 
$(\,(\fr{\tau_J}{\sigma_J})_J, (\alpha_j)_j\,) $ in $\catShat$ with base $b$,
there is a fraction $\fr{\tau}{\sigma}:\HH\to\Sigma$ 
with a 2-cell $\beta_J: \fr{\tau_J}{\sigma_J} \To \fr{\tau}{\sigma}\circ h_J$
for each object $J$ in $\grJ$
such that $\beta_K\circ \alpha_j = \beta_J$ for each arrow $j:J\to K$ in $\grJ$, 
and such a $\fr{\tau}{\sigma}$ is unique up to an in vertible 2-cell. 
Then $\fr{\tau}{\sigma}$ is called ``the'' \emph{lax cotuple} of the $\fr{\tau_J}{\sigma_J}$'s.
This is illustrated below:
$$ 
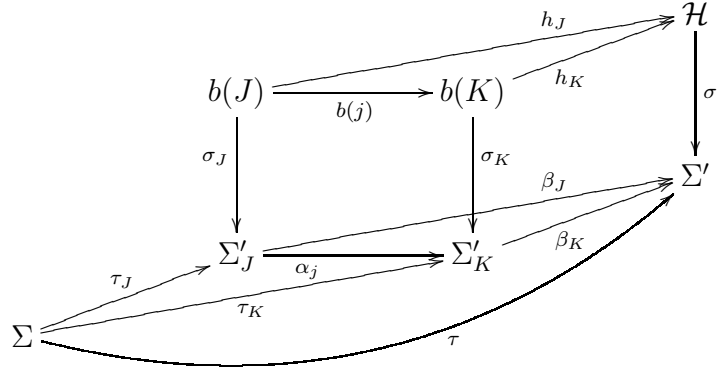
\xymatrix@R=1.2pc@C=5pc{
&&& \HH \ar[dd]^{\sigma} \\
& b(J) \ar[dd]_(.4){\sigma_J} \ar[rru]^(.7){h_J} \ar[r]_{b(j)} &
   b(K) \ar[dd]^(.4){\sigma_K} \ar[ru]_(.4){h_K} & \\
&&& \Sigma' \\
&  \Sigma'_J \ar[rru]^(.7){\beta_J} \ar[r]_(.3){\alpha_j}  & 
   \Sigma'_K \ar[ru]_(.4){\beta_K}  & \\
\Sigma \ar[ru]^{\tau_J} \ar[rru]_{\tau_K} \ar@/_7ex/[rrruu]_(.6){\tau} &&& \\
}$$
It can be proved by diagram-chasing that 
the inclusion functor $\funI:\catS\to\catShat$. 
maps colimits in $\catS$ to lax colimits in $\catShat$.

In any category $\catA$, two objects $X$ and $X'$ are \emph{connected}
if they are related by a chain of morphisms.
Since every functor preserves the connectivity relation,
the \emph{connectivity functor} $\funC:\Cat\to\Set$
maps each category to its set of connected components 
and each functor to the induced map. 

The bicategory $\catShat$ gives rise to the 
\emph{category of classes of fractions} $\catSsim$ 
by identifying the connected fractions.
The \emph{quotient functor} $\funQ:\catShat\to\catSsim$ 
is the identity on objects and maps each morphism to its connectivity class.
In addition, it maps lax colimits to colimits.
Given two specifications $\HH$ and $\Sigma$, 
the set of classes of fractions from $\HH$ to $\Sigma$ can be seen 
from two points of view, since there is a bijection, 
natural in $\HH$ and $\Sigma$ (both in $\catShat$):
\begin{equation}
\label{eqn:ded-Shat-Ssim}
  \funC(\catShat(\HH,\Sigma)) \iso \catSsim(\funQ\HH,\funQ\Sigma)
\end{equation}

The \emph{localization} of a category $\catA$ with respect to 
a set of morphisms $M$ of $\catA$ 
is the functor $\funL_M:\catA\to\catA[M^{-1}]$ 
that consists of adding inverses to the morphisms in~$M$ \cite{GZ67}.
It is easy to check that the functor $\funL=\funQ\circ\funI:\catS\to\catSsim$ 
is the localization of $\catS$ with respect to the set of $\funF$-isomorphisms.
In addition, $\funL$ preserves colimits.
\begin{remark}
Let us emphasize that 
the (bi)categories $\catS$, $\catShat$, $\catSsim$ have the same objects 
(the specifications), which are preserved by the functors 
$\funI$, $\funQ$, $\funL$. 
In addition, the composition of morphisms in $\catSsim$ 
is obtained from the composition of morphisms in $\catShat$,
which itself makes use of composition and pushouts in $\catS$.
\end{remark}

If $\fr{\tau_1}{\sigma_1}$ and $\fr{\tau_2}{\sigma_2}$ are connected, 
then clearly 
$\funF\tau_1^{-1}\circ \funF\sigma_1=\funF\tau_2^{-1}\circ \funF\sigma_2$.
So, a functor $\funFsim:\catSsim\to\catT$ is defined by
$\funFsim(\Sigma)=\funF(\Sigma)$ for each object $\Sigma$ in $\catS$  
and $\funFsim(\funQ(\fr{\tau}{\sigma}))=\funF\tau^{-1}\circ \funF\sigma$
for each fraction $\fr{\tau}{\sigma}$.
It is such that $\funFsim\circ\funL=\funF$.
The functor $\funUsim:\catT\to\catSsim$ is defined as $\funUsim=\funL\circ\funU$.
The next result comes from \cite{GZ67}.
\begin{theorem}
\label{thm:ded-loc}
Let $\funF\dashv\funU:\catS\adjto\catT$ be an adjunction 
where the category $\catS$ is cocomplete. 
Then $\funU$ is full and faithful if and only if $\funFsim$ and $\funUsim$ 
form an equivalence $\funFsim\sim\funUsim:\catSsim\adjto\catT$. 
\end{theorem}
Hence, every logical adjunction $\funF\dashv\funU:\catS\adjto\catT$ 
gives rise to an equivalence $\funFsim\sim\funUsim:\catSsim\adjto\catT$.
It follows that every theory $\Theta$ is isomorphic to $\funFsim\HH$ 
for some specification $\HH$, and that there is a bijection, 
natural in $\Sigma$ and $\HH$ (both in $\catSsim$):
\begin{equation}
\label{eqn:ded-Ssim-T}
  \catSsim(\HH,\Sigma) \iso \catT(\funFsim\HH,\funFsim\Sigma)
\end{equation}
The functors $\funFhat:\catShat\to\catT$ and $\funUhat:\catT\to\catShat$
are now defined by $\funFhat=\funFsim\circ\funQ$ and $\funUhat=\funI\circ\funU$, 
so that $\funQ\circ\funUhat=\funL\circ\funU=\funUsim$.
From the natural bijections (\ref{eqn:ded-Shat-Ssim}) and (\ref{eqn:ded-Ssim-T}),
we get the following bijection,
natural in $\HH$ and $\Sigma$ (both in $\catShat$):
\begin{equation}
\label{eqn:ded-Shat-T}
  \funC(\catShat(\HH,\Sigma)) \iso \catT(\funFhat\HH,\funQ\funFhat\Sigma)
\end{equation}

\subsection{Inference steps}
\label{subsec:ded-step}

Given a logical adjunction $\funF\dashv\funU:\catS\adjto\catT$, 
we have defined the bicategory $\catShat$ of fractions, 
with an inclusion functor $\funI:\catS\to\catShat$ 
and a quotient functor $\funQ:\catShat\to\catSsim$, 
such that the localization of $\catS$ with respect to the $\funF$-isomorphisms 
is $\funL=\funQ\circ\funI:\catS\to\catSsim$. 

\begin{definition}
\label{defi:ded-entail}
An \emph{entailment} is a $\funF$-isomorphism $\tau:\Sigma\ento\Sigma'$,
and an \emph{instance of $\Sigma$ in $\Sigma_1$} is a fraction 
$\fr{\tau}{\sigma}:\Sigma\to\Sigma'_1\lento\Sigma_1$ 
(also written $\fr{\tau}{\sigma}:\Sigma\to\Sigma_1$, in~$\catShat$). 
Let us consider a fraction $\rho:\CC\to\HH'\lento\HH$,
called an \emph{inference rule}
with \emph{hypothesis} $\HH$ and \emph{conclusion} $\CC$.
Given a specification $\Sigma$, 
the \emph{inference step in $\Sigma$ along $\rho$} 
is the functor $\catShat(\rho,\Sigma):\catShat(\HH,\Sigma)\to\catShat(\CC,\Sigma)$
of composition on the right with $\rho$.
\end{definition}
\begin{remark}
Each entailment $\tau$ gives rise to an isomorphism $\funF\tau$
and to a bijection $\Mod(\tau,\Theta)$ for each theory $\Theta$,
which proves the \emph{soundness} of the logical adjunction. 
Each instance $\fr{\tau}{\sigma}$ of $\Sigma$ in $\Sigma'$
gives rise to a morphism 
$\funF(\fr{\tau}{\sigma}):\funF\Sigma\to\funF\Sigma'$ 
and to a function  
$\Mod(\fr{\tau}{\sigma},\Theta):\Mod(\Sigma',\Theta)\to\Mod(\Sigma,\Theta)$
for each $\Theta$.
\end{remark}
The inference step in $\Sigma$ along $\rho$
maps each instance $\kappa$ of $\HH$ in $\Sigma$
to the instance $\gamma=\kappa\circ\rho$ of $\CC$ in $\Sigma$.
The composition is performed in $\catShat$, 
which means that it requires a pushout in $\catS$. 
This is illustrated below. 
We have chosen an illustration that is different from 
the illustration of composition of fractions in section~\ref{subsec:ded-frac},
because it better reflects the semantics of inference: 
the top line is made of the rule, with its hypothesis on the left, 
the bottom line is made of the given specification $\Sigma$ 
and its entailments; the square on the left is a pushout,  
and the diagram is commutative.
The numerator and denominator of $\rho$
are denoted $\sigma\rho$ and $\tau\rho$,
and similarly for $\kappa$ and $\gamma$.
$$ \begin{array}{lll} 
\mbox{in $\catShat:$} & \qquad &  \mbox{in $\catS:$} \\
\quad \xymatrix@C=3.5pc{
\HH \ar[d]_{\kappa} & \CC  \ar[l]_{\rho} \ar[d]^{\gamma} \\
\Sigma \ar@{=}[r] & \Sigma \\
} & &
\quad \xymatrix@C=3.5pc{
& \HH \ar[r]_{\tau\rho} \ar[d]_{\sigma\kappa} & \HH' \ar@<-1ex>@{-->}[l] \ar[d] & 
  \CC \ar[l]_{\sigma\rho} \ar[d]^{\sigma\gamma} \\
\Sigma \ar[r]_{\tau\kappa} \ar@<-2ex>@/_4ex/[rrr]_(.7){\tau\gamma} & 
  \Sigma_\kappa \ar[r]  \ar@<-1ex>@{-->}[l]  & 
  \Sigma_\gamma \ar@{=}[r]  \ar@<-1ex>@{-->}[l] & 
  \Sigma_\gamma \ar@<1ex>@/^4ex/@{-->}[lll]\\
} \\
\end{array}$$
\begin{example}[Modus ponens]
\label{exa:ded-mop}
Let us consider the logical adjunction for which 
a specification is a pair $\Sigma=\lr{\Sigma_F,\Sigma_P}$ 
made of a set $\Sigma_F$ of \emph{formulas} with a partial binary operator ``$\To$'' 
and a subset $\Sigma_P\subseteq\Sigma_F$ of \emph{provable formulas},
and a theory $\Theta$ is a specification that satisfies two properties: 
when $A$ and $B$ are formulas then $A\To B$ is a formula, 
and when $A\To B$ and $A$ are provable then $B$ also is provable.
The second property corresponds to
the \emph{modus ponens} inference rule $\CC_P\to\HH'_P\lento\HH_P$ 
for building provable formulas,
where $\HH_P=\lr{ \{A,B,A\To B\},\{A,A\To B\} }$, 
$\HH'_P=\lr{ \{A,B,A\To B\},\{A,A\To B,B\} }$, 
$\CC_P=\lr{ \{C\},\{C\} }$,
the entailment $\HH_P\ento\HH'_P$ is the inclusion
and the morphism $\CC_P\to\HH'_P$ maps $C$ to $B$.
Classically, only the provable formulas of $\HH_P$ and 
of the image of $\CC_P$ in $\HH'_P$ are mentioned, 
and the modus ponens rule is written $\frac{A\;\;A\To B}{B}$. 
The hypothesis $\HH_P=\lr{ \{A,B,A\To B\},\{A,A\To B\} }$ 
contains two provable formulas.
It is made of two simpler hypothesis 
$\HH_1=\lr{ \{A\},\{A\} }$ and $\HH_2=\lr{ \{A,B,A\To B\},\{A\To B\} }$.
More precisely, $\HH_P$ is the colimit of the diagram $\HH_1 \lto \HH_0 \to \HH_2 $, 
where $\HH_0=\lr{ \{A\},\emptyset }$ and the morphisms are the inclusions.
So, each instance of $\HH_P$ can be built as a lax cotuple of instances. 
\end{example}

\subsection{Inference process} 
\label{subsec:ded-proc}

\begin{definition}
\label{defi:ded-proof}
With respect to a set $\logR$ of inference rules, 
a \emph{proof} is a fraction in the sub cocomplete bicategory of $\catShat$
which is generated by $\logR$.
\end{definition}
In this section, for simplicity, it is assumed that the proofs are all the fractions;
this assumption is discussed in section~\ref{sec:syn}.
The inference process is defined now by allowing every ingredient 
(either $\HH$, $\rho$, $\Sigma$, or $\kappa$)
of the inference step to vary in the relevant (bi)category.
When $\catC$ and $\catD$ are bicategories, 
we define a \emph{functor} from $\catC$ to $\catD$
as a homomorphism in \cite{Le98}. 
Essentially, a functor $\funG:\catC\to\catD$
maps objets to objects, morphisms to morphisms, 2-cells to 2-cells, 
it preserves composites of 2-cells, 
and it preserves composites of morphisms only up to an invertible 2-cell.
A \emph{contravariant functor}  $\funG:\catC\opto\catD$
is contravariant on morphisms and covariant on 2-cells.

The inference steps in $\Sigma$ can be composed, 
in two slightly different ways.
Let $\rho_1:\CC_1\to\HH_1$ and $\rho_2:\CC_2\to\HH_2$
with $\CC_1=\HH_2$ in $\catShat$, 
and let $\kappa$ be an instance of $\HH_1$ in $\Sigma$.
On the one hand, $\kappa$ is mapped 
by $\catShat(\rho_2,\Sigma)\circ\catShat(\rho_1,\Sigma)$
to $(\kappa\circ\rho_1)\circ\rho_2$.
On the other hand, $\kappa$ is mapped  
by $\catShat(\rho_1\circ\rho_2,\Sigma)$ 
to $\kappa\circ(\rho_1\circ\rho_2)$.
These are two instances of $\CC_2$ in $\Sigma$ related by an invertible 2-cell. 
The following definition corresponds to the second point of view.
\begin{definition}
\label{defi:ded-proc-in}
The \emph{inference process in $\Sigma$} 
is the contravariant functor $\catShat(-,\Sigma):\catShat\opto\Cat$. 
\end{definition}
The contravariant functor $\catShat(-,\Sigma)$ 
maps each specification $\HH$ to the category $\catShat(\HH,\Sigma)$ 
and each proof $\rho:\CC\to\HH$  to the functor $\catShat(\rho,\Sigma)$
of composition on the right with $\rho$, as for primitive inference rules:
this means that a proof is seen as a \emph{derived inference rule} $\rho$.

Now $\Sigma$ itself may vary. 
Let ${\Cat}^{\catShat,\op}$ denote 
the 2-category of contravariant functors from $\catShat$ to $\Cat$ 
and $\funY'_\catShat:\catShat\Lto{\Cat}^{\catShat,\op} $ 
the Yoneda covariant functor of $\catShat$, 
which maps each specification $\Sigma$ to the contravariant functor 
$\funY'\Sigma=\catShat(-,\Sigma)$ \cite{Le98}.
\begin{definition}
\label{defi:ded-proc}
The \emph{inference process} 
is the Yoneda functor $\funY'_\catShat \!:\! \catShat\to{\Cat}^{\catShat,\op}\!$. 
\end{definition}

\section{Syntax} 
\label{sec:syn}

\subsection{Propagators} 
\label{subsec:syn-lsk}

In order to define a \emph{syntax} for some logical adjunctions,
we use limit sketches. 
Among all kinds of sketches, the limit sketches play a very special
role in this paper, since they are used to define the ``logic for logics'', 
\ie, the \emph{meta logic} for defining all diagrammatic logics,
as explained below. 
As in example~\ref{exa:mod-lim}, 
a \emph{limit sketch} $\skE$ is a graph where 
some points $X$ 
    have a \emph{(potential) identity} $\id_X:X\to X$,
some consecutive arrows $f:X\to Y$, $g:Y\to Z$ 
    have a \emph{(potential) composite} $g\circ f:X\to Z$,
and some diagrams $b:\grJ\to\skE$ 
    have a \emph{(potential) limit} $(p_J:\Lim(b)\to X_J)_J$ with base $b$.
A morphism of limit sketches is a morphism of graphs 
which preserves all potential features, 
which gives rise to the category $\LSk$ of limit sketches.

\begin{definition}
\label{defi:syn-real}
A \emph{realization} of a limit sketch $\skE$ in any category $\catC$ 
is a morphism of graphs which maps the potential features in $\skE$
to actual features in $\catC$. 
A \emph{morphism of realizations of $\skE$ in $\catC$} 
is a natural transformation.
This yields the category $\Rea(\skE,\catC)$ of realizations of $\skE$ in $\catC$.
\end{definition}
The realizations of $\skE$ could be called its ``loose models''. 
The category $\Rea(\skE)=\Rea(\skE,\Set)$ is cocomplete. 
When a category $\catA$ is equivalent to $\Rea(\skE)$, 
we say that $\skE$ is \emph{a sketch for}  $\catA$.

\begin{definition}
\label{defi:syn-prop}
A  \emph{propagator} $P:\skE_S\to\skE_T$ is a morphism of limit sketches.
A \emph{morphism of propagators} $\ell:P\to P'$, 
where $P:\skE_S\to\skE_T$ and $P':\skE'_S\to\skE'_T$,
is made of two propagators
$\ell_S:\skE_S\to\skE'_S$ and $\ell_T:\skE_T\to\skE'_T$
such that $\ell_T\circ P = P'\circ\ell_S$.
This yields the category of propagators. 
\end{definition}

For each propagator $P:\skE_S\to\skE_T$, 
the \emph{underlying functor} $\funU_P=\Rea(P):\Rea(\skE_T)\to\Rea(\skE_S)$
is neither full nor faithful, in general.
A fundamental result about limit sketches \cite{Eh68}
is that  this underlying functor has a left adjoint
$\funF_P:\Rea(\skE_S)\to\Rea(\skE_T)$,
called the \emph{freely generating functor}.
So, every propagator gives rise to an adjunction 
$\funF_P\dashv\funU_P:\Rea(\skE_S)\adjto\Rea(\skE_T)$ 
where $\Rea(\skE_S)$ is cocomplete. 

\begin{definition}
\label{defi:syn-equiv}
A propagator $P$ is an \emph{equivalence of limit sketches}
when the adjunction $\funF_P\dashv\funU_P$ is an equivalence of categories. 
A morphism of propagators $\ell:P\to P'$ is an \emph{equivalence}
when both $\ell_S$ and $\ell_T$ are equivalences of limit sketches.
\end{definition}

The Yoneda contravariant functor can be generalized to limit sketches, 
as follows \cite{LD01,Du03}. 
Let $\skE$ be a limit sketch and $\Proto(\skE)$ its \emph{prototype},
\ie, the category generated by $\skE$ in such a way that 
all potential features in $\skE$ become actual features in $\Proto(\skE)$.
The Yoneda contravariant functor $\funY_{\Proto(\skE)}$
is such that its restriction to $\skE$ forms a contravariant realization 
of $\skE$ with values in $\Rea(\skE)$.  
This is the \emph{Yoneda contravariant realization of $\skE$}, denoted $\funY_\skE$.
The \emph{density property} of $\funY_\skE$ states that 
every realization $\Sigma$ of $\skE$ is the colimit of a diagram 
in the image of $\funY_\skE$. 
When $P:\skE_S\to\skE_T$ is a propagator, 
the Yoneda contravariant realizations $\funY_S$ and $\funY_T$ of $\skE_S$ and $\skE_T$,
respectively, are such that there is a natural isomorphism 
$\funF_P\circ\funY_S \iso \funY_T\circ P$.

\subsection{Logical propagators} 
\label{subsec:syn-prop}

\begin{definition}
\label{defi:syn-logic}
A \emph{logical propagator} is a propagator $P:\skE_S\to\skE_T$ such that the 
underlying functor $\funU_P$ is full and faithful.
\end{definition}
Since the category $\Rea(\skE_S)$ is always cocomplete, 
every logical propagator $P:\skE_S\to\skE_T$ gives rise to a logical adjunction
$\funF_P\dashv\funU_P:\Rea(\skE_S)\adjto\Rea(\skE_T)$.

For instance, for each set of arrows $A$ of~$\skE$,
the \emph{localizer} of~$\skE$ with respect to $A$ is the propagator with source $\skE$ 
which, for each $a:E'\to E$ in $A$, adds an arrow $a^{-1}:E\to E'$
and the composites $a\circ a^{-1}=\id_E$ and $a^{-1}\circ a=\id_{E'}$.
It is is easy to check that when $P$ is a localizer then it is a logical propagator 
and the functor $F_P$ is a localization.
In addition, a morphism of propagators $\ell:P\to P'$ where 
$P$ is logical is characterized by $\ell_S:\skE_S\to\skE'_S$
such that each $P$-entailment in $\skE_S$ is mapped to 
a $P'$-entailment in $\skE'_S$.

As an instance of a (generally) non-logical propagator,
let us say that a propagator $P:\skE\to\skE'$ is a \emph{swelling propagator} 
if it is an inclusion
and if every arrow of $\skE'$ with its source in $\skE$ is an arrow in $\skE$,
every composition of $\skE'$ with its source in $\skE$ is 
a composition in $\skE$,
every limit of $\skE'$ with its vertex in $\skE$ is a limit in $\skE$,
and every other limit of $\skE'$ has at least one projection entirely outside $\skE$
(which means that the vertex of the limit, at least one of its projection, 
and the target of this projection, are outside $\skE$).
Then the freely generating functor $\funF_P$
consists of ``adding nothing'', in the following sense:
let $\Sigma$ be a realization of $\skE$,
then $\funF_P\Sigma$ is such that its restriction to $\skE$
coincides with $\Sigma$ and 
$\funF_P\Sigma(E')=\emptyset$ for each point $E'$ not in $\skE$,
so that $\funF_P\Sigma(e')$ is the unique map with source $\emptyset$ 
for each arrow $e'$ not in $\skE$.
It is clear that $\funF_P$ is full and faithful.

From now on, for simplicity, it is assumed that each limit sketch
has ``enough'' identities: 
either there is an identity for each point, 
or at least there is an identity whenever we need it
(adding identities is an equivalence).
The next result is the \emph{decomposition theorem} from \cite{Du03}.
Such a decomposition is not uniquely determined.
\begin{theorem}
\label{thm:syn-dec}
For each propagator $P:\skE_S\to\skE_T$
there is a swelling propagator $P_S$
and a logical propagator $P'$ such that $P = P' \circ P_S$. 
In addition, there is such a decomposition where $P'$ is composed 
of a localizer followed by an equivalence. 
\end{theorem}
As a typical example, here is a decomposition of a propagator $P$
which adds an arrow between two given points:
$P_S$ is the inclusion and $P'$ maps $t_H$ and $t_C$ 
to $\id_H$ and $\id_C$, respectively.
So, $P'$ is composed of the localizer with respect to $\{t_H,t_C\}$
followed by the equivalence that maps $t_H$, $t_C$ and their inverses 
to identities.
$$\begin{array}{|c|c|c|c|c|}
\cline{1-1}\cline{3-3}\cline{5-5}  
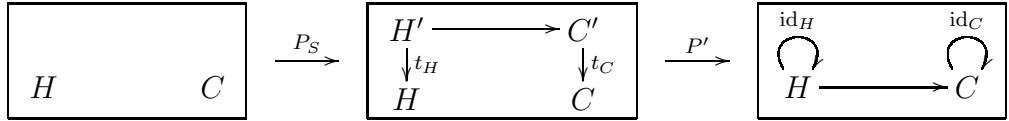
\xymatrix@C=4pc@R=1pc{
  \mbox{ } & \\  H&C\\}
 & \xymatrix@R=.5pc{\mbox{} \\ \mbox{} \ar[r]^{P_S} & \mbox{} \\} & 
\xymatrix@C=4pc@R=1pc{
  H'\ar[d]^{t_H} \ar[r]&C'\ar[d]^{t_C} \\ 
  H &C\\}
 & \xymatrix@R=.5pc{\mbox{} \\ \mbox{}\ar[r]^{P'} & \mbox{} \\} & 
\xymatrix@C=4pc@R=1pc{
  \mbox{ } & \\  H\ar@(ul,ur)^{\id_H} \ar[r]&C\ar@(ul,ur)^{\id_C} \\} 
\\ 
\cline{1-1} \cline{3-3}\cline{5-5}  
\end{array}$$

\subsection{Diagrammatic logics}
\label{subsec:syn-dialog}

Clearly, a propagator that is equivalent to a logical propagator is also logical.
\begin{definition}
\label{defi:syn-dialog}
A \emph{diagrammatic logic} $\logL$ is an equivalence class of logical propagators.
Hence the morphisms of diagrammatic logics are defined from the morphisms of 
propagators, which yields the \emph{category} $\DiaLog$ of diagrammatic logics. 
An \emph{inference system} for a diagrammatic logic $\logL$ is a 
localizer in the class~$\logL$.
\end{definition}
Theorem~\ref{thm:syn-dec} provides a diagrammatic logic
and an inference system for this logic, from any propagator.
Let  $\logL$ be a diagrammatic logic,
and $P$ a chosen logical propagator in the class $\logL$.
The following notions are defined with respect to $\logL$ and $P$.
\begin{definition}
\label{defi:syn-spth}
The \emph{specifications}, \emph{theories} and \emph{models} 
are the specifications, theories and models with respect to 
the logical adjunction $\funF_P\dashv\funU_P$.
When in addition $P$ is an inference system for $\logL$, 
a \emph{syntactic inference rule} is a (right) fraction $r=\frsyn{s}{t}:H\lento H'\to C$
where $s: H'\to C$ is any arrow in $\skE_S$ 
and $t:H'\ento H$ is an arrow in $\skE_S$ such that $P(t)$ is invertible in $\skE_T$.
\end{definition}
So, the image of a syntactic inference rule by the Yoneda contravariant functor of $\skE_S$ 
is an instance with respect to the logical adjunction $\funF_P\dashv\funU_P$,
in coherence with definition~\ref{defi:ded-entail}.

The \emph{type} $\Type(\skE)$ of a limit sketch $\skE$ 
is the complete category generated by $\skE$ in such a way that 
all potential features in $\skE$ become actual features in $\Type(\skE)$.
Each propagator $P:\skE_S\to\skE_T$ gives rise to a limit-preserving functor 
$\Type(P):\Type(\skE_S)\to\Type(\skE_T)$. 
A \emph{syntactic proof} is defined as a syntactic inference rule,
with respect to $\Type(P)$ instead of $P$. 
So, the image of a syntactic proof by the Yoneda contravariant functor 
is a proof with respect to the logical adjunction $\funF_P\dashv\funU_P$,
as in definition~\ref{defi:ded-proof}.
The density property of the Yoneda contravariant functor shows that every proof
with respect to $\funF_P\dashv\funU_P$ is isomorphic to the image of a syntactic proof,
which justifies the assumption that ``the proofs are all the fractions''
in section~\ref{subsec:ded-proc}.
\begin{remark}
Our definition of diagrammatic logics and their deduction processes 
provides a new point of view about the notion of
inference process in a specification $\Sigma$ (definition~\ref{defi:ded-proc-in}),
which now can be seen as \emph{a realization of $\skE_S$ with values in $\Cat$}.
Indeed, the Yoneda contravariant realization $\funY_S$ of $\skE_S$
takes its values in the category $\catS=\Rea(\skE_S)$,
so that it can be composed with the inference process in $\Sigma$,
\ie, with the contravariant functor $\catShat(-,\Sigma):\catShat\opto\Cat$,
which maps colimits to limits. This gives rise to 
the realization $\catShat(-,\Sigma)\circ\funY_S$ of $\skE_S$ with values in $\Cat$.
\end{remark}
\begin{example}[Equational logic: syntax]
\label{exa:syn-equ}
As reminded in example~\ref{exa:mod-equ}, 
the equational logic can be defined from the full reflection of $\FpSk$ in $\FpCat$. 
This logical adjunction comes from a logical propagator 
$P_{\eq}:\skE_{\fpsk}\to\skE_{\fpcat}$, 
that defines the \emph{diagrammatic equational logic} $\logL_{\eq}$.
This propagator can be obtained from a propagator 
$P_{\fpcat}:\skE_{\gr}\to\skE_{\fpcat}$ for the reflection of $\Gr$ in $\FpCat$,
by a decomposition satisfying the properties of theorem~\ref{thm:syn-dec}.
This is also the case for the limit logic in example~\ref{exa:mod-lim}
and for other kinds of doctrines. 
\end{example}

\subsection{Finiteness issues}
\label{subsec:syn-fin}

No finiteness condition has been assumed until now.
However, a syntax is used for writing things down 
with a finite number of symbols. So that we have to check some finiteness 
properties, in order to ensure that an inference system 
for a diagrammatic logic does define a syntax. 
This issue is outlined now. 

Let $P:\skE_S\to\skE_T$ be an inference system, 
and let us assume that the sketches $\skE_S$ and $\skE_T$ are finite
(the categories $\catS=\Rea(\skE_S)$ and $\catT=\Rea(\skE_T)$ are usually infinite).
Then, each specification $\Sigma$ is defined from the finite number of sets $\Sigma(E)$,
for all points $E$ of $\skE_S$,
and there is a finite number of elementary inference rules.

In addition, let us assume that $\Sigma$ is finite, 
in the sense that each set $\Sigma(E)$ is finite
(the generated theory $\funF\Sigma$ is usually infinite).
The inference process builds new specifications $\Sigma'$,
which are entailed from $\Sigma$. It has to be checked that 
these specifications also can be assumed finite.  
Each $\Sigma'$ is built as the vertex of a pushout in $\catS$,
so that is is finite as soon as  
\emph{the finite colimits in $\catS$ preserve finiteness},
which means that the colimit of a finite base made of finite specifications 
is a finite specification.
This need not be true, in general.
This issue can be solved thanks to an assumption about the \emph{acyclicity}
of sketches, as explained below. This assumption is presumably rather strong. 

Usually a \emph{cycle} in a graph is a loop, distinct from an identity,
in the generated category. 
Let us define a \emph{cycle} in a limit sketch as loop, distinct from an identity,
in the generated type. 
When a limit sketch $\skE$ is acyclic, 
then the finite colimits in $\Rea(\skE)$ preserve finiteness. 
Let us assume that $P:\skE_S\to\skE_T$ is built thanks to theorem~\ref{thm:syn-dec},
from a propagator $P_0:\skE_{S,0}\to\skE_{T,0}$,
with $\skE_{S,0}$ acyclic (like for instance $\skE_{\gr}$). 
Then the construction in the proof of theorem~\ref{thm:syn-dec} 
can be modified in such a way that $\skE_S$ also is acyclic:
basically, every arrow $f:X\to Y$ in $\skE_{T,0}$ 
gives rise to a span $X\lupto{c} X' \upto{f'} Y$ in $\skE_S$,
which is mapped to $X\lupto{\id} X \upto{f} Y$ in $\skE_T$.
The idea is that $c$ stands for an injection 
and $f'$ for a partial version of $f$
(it can be added that $c$ is a potential monomorphism).
Then the finite colimits in $\catS$ preserve finiteness,
which is the required property.

\section{Applications}
\label{sec:exa}

\subsection{Decorations}
\label{subsec:exa-dec}

The notion of \emph{decoration}, as defined below, 
can be used for studying the semantics of computer languages. 
For instance, for dealing with multivariate functions in imperative programming,
as explained below and in \cite{DDR07}.
It may also be used for formalizing the mechanism of exceptions \cite{DR06}.
The idea of \emph{decoration}
is based upon a span of diagrammatic logics:
\vspace{-5pt} 
$$\xymatrix@C=4pc{
\logL_{\simp} & \logL_{\deco} \ar[l]_{\ell_{\simp}} \ar[r]^{\ell_{\expl}} & \logL_{\expl} \\ 
} \vspace{-2pt} $$
where the three logics $\logL_{\deco}$, $\logL_{\simp}$ and $\logL_{\expl}$
are called respectively the \emph{decorated}, \emph{simplified} 
and \emph{explicit} logics,
and the morphisms $\ell_{\simp}$ and $\ell_{\expl}$ are 
the \emph{simplification}  and \emph{explicitation} morphisms. 
The subscripts in the notations are simplified, for instance 
the adjunctions (either on specifications or on theories)
with respect to $\ell_{\simp}$ 
are denoted $\funF_{\simp}\dashv\funU_{\simp}$, and so on.
It is assumed that both freely generating functors on specifications 
$\funF_{\simp}$ and $\funF_{\expl}$ are easy to compute. 
The idea is that the logics $\logL_{\simp}$ and $\logL_{\expl}$ are well-known, 
while the decorated logic $\logL_{\deco}$ is not.
The morphisms $\ell_{\simp}$ and $\ell_{\expl}$
are used for building proofs and models, respectively, 
for any given decorated specification $\Sigma_{\deco}$.

On the models side, 
it is assumed that the set of intended models of $\Sigma_{\deco}$
is $\Mod_{\expl}(\Sigma_{\expl},\Theta_{\expl})$ for 
some given explicit theory $\Theta_{\expl}$,
where $\Sigma_{\expl}=\funF_{\expl}\Sigma_{\deco}$.
Then, according to proposition~\ref{prop:mod-mod},
the set of intended models of $\Sigma_{\deco}$ can be identified with 
$\Mod_{\deco}(\Sigma_{\deco},\Theta_{\deco}) $
where $\Theta_{\deco}=\funU_{\expl}\Theta_{\expl}$.
This ensures the soundness of the intended models of $\Sigma_{\deco}$ 
with respect to the proofs in the decorated logic.

On the proofs side, a decorated proof $p_{\deco}$ is mapped by $\ell_{\simp,T}$
to a simplified proof $p_{\simp}$. 
This property provides a method for building decorated proofs in two steps:
first a simplified proof $p_{\simp}$ is built in the well-known simplified logic, 
then, if possible, a decorated proof $p_{\deco}$ is built such that
$p_{\simp}=\ell_{\simp,T}p_{\deco}$. 

\subsection{Multivariate functions in imperative programming}
\label{subsec:exa-prod}

Multivariate functions in functional (effect-free) programming 
can be formalized via categorical products: 
a term $f(t_1,t_2)$ is composed of the pair $\lr{t_1,t_2}$ 
followed by the bivariate function $f$, so that $t_1$ and $t_2$
play symmetric roles. This cannot be done in imperative programming, 
where the value of $f(t_1,t_2)$ 
may depend on the order of evaluation of $t_1$ and $t_2$.
A major contribution of \cite{DDR07} is the definition of 
the \emph{sequential product} of morphisms in an effect category,
for formalizing ``first $t_1$, then $t_2$''.
Then a \emph{cartesian effect category} is an effect category
with sequential products, it provides a 
semantics for computational languages with effects.
This is shortly reminded below,
by looking at the diagrammatic logics that are involved.

The \emph{simplified logic} $\logL_{\simp}$ is the equational logic $\logL_{\eq}$, 
defined as the class of the logical propagator 
$P_{\eq}$ in example~\ref{exa:syn-equ}.

Let us define the \emph{decorated logic} $\logL_{\deco}$. 
Let $\catV$ be  category, 
a \emph{(strict) effect category} extending $\catV$
is a category $\catC$ such that $\catV$ is a wide subcategory of $\catC$
(the morphisms in $\catV$ are called \emph{pure})
and $\catC$ is endowed  with a \emph{semi-congruence}~$\eqwe$,
\ie, a reflexive and transitive relation between parallel morphisms in $\catC$ 
which satisfies the substitution property 
and only a ``pure'' version of the replacement property: 
if $g_1\eqwe g_2:Y\to Z$ then 
$g_1\circ f\eqwe g_2\circ f$ \emph{for all $f:X\to Y$ in $\catC$}
and $v\circ g_1 \eqwe v\circ g_2$ \emph{for all $v:Z\to W$ in $\catV$}.
In a cartesian category, the product $t_1\times t_2$ of two morphisms 
is the unique morphism 
such that $q_1\circ(t_1\times t_2)=t_1\circ p_1$
and $q_2\circ(t_1\times t_2)=t_2\circ p_2$,
where the $p_i$'s and $q_i$'s are the relevant projections.
In a cartesian effect category, such a product is defined for pure morphisms.
When $t_2$ is pure but $t_1$ is not, the \emph{semi-product} 
$t_1\times t_2$ is characterized by 
$q_1\circ(t_1\times t_2)=t_1\circ p_1$
and \emph{only} $q_2\circ(t_1\times t_2)\eqwe t_2\circ p_2$.
Then the \emph{sequential product} of two morphisms $t_1$ and $t_2$, 
when maybe neither is pure, 
is the composition of the semi-products $(t_1\times\id)$ and $(\id\times t_2)$. 
For dealing with the side-effects due to modifications of a global state,
the relation $f\eqwe g$  
means that the functions $f$ an $g$ return the same result, 
but they may modify the state in two different ways
(so that, in this case, $\eqwe$ is interpreted as an equivalence relation). 
Like $P_{\eq}$ is obtained from a decomposition of $P_{\fpcat}$ 
in example~\ref{exa:syn-equ},
the logical propagator $P_{\effeq}:\skE_{\fpeffsk}\to\skE_{\fpeffcat}$
is obtained from a decomposition of 
a propagator $P_{\fpeffcat}:\skE_{\effgr}\to\skE_{\fpeffcat}$, 
where $\skE_{\effgr}$ is the limit sketch for \emph{effect graphs}:
\vspace{-7pt} 
  $$ \xymatrix@C=5pc{
  \mathtt{Point} & 
  \mathtt{Arrow} \ar@/_/[l]_{\mathtt{source}} \ar@/^/[l]^{\mathtt{target}} 
  & \; \; \mathtt{PureArrow} \ar[l]_{\mathtt{inj}} \\ }
\vspace{-5pt} $$
where $\mathtt{PureArrow}$ and $\mathtt{inj}$ stand respectively 
for the set of \emph{pure arrows} and for the conversion
(it can be added that $\mathtt{inj}$ is a potential monomorphism). 
This gives rise to the \emph{diagrammatic equational logic with effects}
$\logL_{\deco}$.

The \emph{simplification morphism} $\ell_{\simp}:\logL_{\deco}\to\logL_{\simp}$
maps $\mathtt{PureArrow}$ to $\mathtt{Arrow}$ 
and $\mathtt{inj}$ to $\mathtt{id}_{\mathtt{Arrow}}$,
which means that it blurs the distinction between pure and non-pure morphisms.
Similarly, $\ell_{\simp}$ maps the semi-congruence to the equality.
It follows that $\ell_{\simp}$ maps sequential products to ordinary products.
So, each proof in $\logL_{\deco}$ is mapped to a proof in $\logL_{\simp}$.
This property is used in the appendix of \cite{DDR07} for building proofs 
in equational logic with effects, by decorating proofs in equational logic.
The intended models of the decorated specifications
are \emph{not} preserved by the simplification morphism.

The \emph{explicit logic} $\logL_{\expl}$ is the  \emph{pointed equational logic},
made of the equational logic together with a distinguished sort $S$
of \emph{states}. The morphisms 
of pointed equational specifications (\resp theories) must preserve $S$.
It is easy to build $\logL_{\expl}$ from the equational logic $\logL_{\eq}$. 

The \emph{explicitation morphism} $\ell_{\expl}:\logL_{\deco}\to\logL_{\expl}$
is based on the idea that a morphism $f:X\to Y$ in a decorated specification 
is mapped to a morphism $f:S\times X\to S\times Y$, 
and that when $f$ is pure then $f=\id_S\times f_0$ for some $f_0:X\to Y$.
This informal description corresponds to the formal description 
of $\ell_{\expl}$ via the Yoneda contravariant realizations,
using the natural isomorphism $\funF_P\circ\funY_S \iso \funY_T\circ P$
(for every propagator $P$).
The image of $\skE_{\effgr}$ by the Yoneda contravariant realization 
of $\skE_{\effgr}$ in the category of effect graphs is as follows
(pure morphisms are represented as dashed arrows):
\vspace{-2pt} 
$$ \begin{array}{|c|c|c|c|c|}
\cline{1-1}\cline{3-3}\cline{5-5} 
\xymatrix{X\\} &
\xymatrix@R=-0.3pc@C=4pc{
  \mbox{}\ar@/^1ex/[r]^{X\mapsto X} & \mbox{}\\  
  \mbox{}\ar@/_1ex/[r]_{X\mapsto Y} & \mbox{} \\} &
\xymatrix{X\ar[r]^{f}&Y\\} & 
\xymatrix@C=4pc{
  \mbox{} \ar[r]  & \mbox{}\\ } & 
\xymatrix{X\ar@{-->}[r]^{f}&Y\\} \\
\cline{1-1}\cline{3-3}\cline{5-5} 
\end{array}\vspace{-2pt} $$
This is mapped by $\funF_{\expl}$ to the following diagram in the 
category of pointed finite product sketches (the vertical arrows are the projections): 
\vspace{-2pt} 
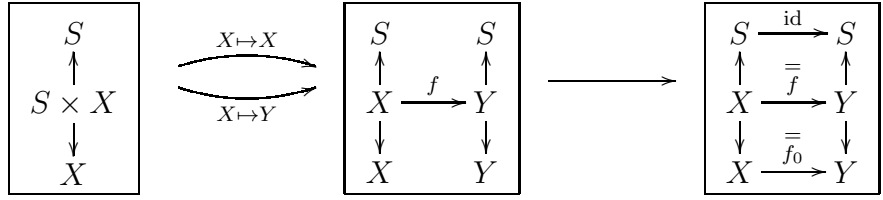
$$ \begin{array}{|c|c|c|c|c|}
\cline{1-1}\cline{3-3} \cline{5-5}
\xymatrix@R=1pc{S\\ S\times X\ar[u]\ar[d]\\ X\\} &
\xymatrix@R=0pc@C=4pc{
  \mbox{ \rule{0pt}{15pt} } \\
  \mbox{}\ar@/^1ex/[r]^{X\mapsto X} & \mbox{}\\  
  \mbox{}\ar@/_1ex/[r]_{X\mapsto Y} & \mbox{} \\} &
\xymatrix@R=1pc{S&S \\ X\ar[u]\ar[d]\ar[r]^{f}&Y\ar[u]\ar[d]\\ X&Y\\} &
\xymatrix@R=1pc@C=4pc{
  \mbox{} \\ 
  \mbox{}  \ar[r] & \mbox{}\\ } & 
\xymatrix@R=1pc{S\ar[r]^{\id}&S \\ 
  X\ar[u]\ar[d]\ar[r]^{f} \ar@{}[ur]|{=} \ar@{}[dr]|{=}&Y\ar[u]\ar[d]\\ X\ar[r]^{f_0}&Y\\} \\
\cline{1-1}\cline{3-3} \cline{5-5}
\end{array}\vspace{-2pt} $$
For a fixed set of states $\setS$,
the category of sets together with $\setS$ forms a 
theory $\Set_{\setS}$ with respect to the pointed equational logic.
The intended models of a decorated specification $\Sigma_{\deco}$
can be defined as the models of the explicit specification 
$\Sigma_{\expl}=\funF_{\expl}\Sigma_{\deco}$ 
with values in $\Set_{\setS}$.
As explained in section~\ref{subsec:exa-dec},
this ensures the soundness of the intended models of $\Sigma_{\deco}$ 
with respect to the proofs in the decorated logic.




\end{document}